\documentclass[11pt,twoside]{article}
\usepackage{asp2010}

\resetcounters

\begin{document}

\title{Molecfit: A Package for Telluric Absorption Correction}
\author{W. Kausch$^1$, S. Noll$^1$, A. Smette$^2$, S. Kimeswenger$^{3,1}$, H. Horst$^4$, H. Sana$^5$, A. Jones$^1$, M. Barden$^1$, C. Szyszka$^1$, and J. Vinther$^6$ }
\affil{$^1$Institute for Astro- and Particle Physics, University of Innsbruck, A-6020 Innsbruck}
\affil{$^2$European Southern Observatory, Santiago, Chile}
\affil{$^3$Instituto de Astronom\'ia, Universidad Cat\'olica del Norte, Antofagasta, Chile}
\affil{$^4$EOS Electro Optical Systems, M\"unchen, Germany}
\affil{$^5$Astronomical Institute Anton Pannekoek, University of Amsterdam, The Netherlands}
\affil{$^6$European Southern Observatory, Garching, Germany}

\begin{abstract}
Correcting for the sky signature usually requires supplementary calibration data which are very expensive in terms of telescope time. In addition, the scheduling flexibility is restricted as these data have to be taken usually directly before/after the science observations due to the high variability of the telluric absorption which depends on the state and the chemical composition of the atmosphere at the time of observations. Therefore, a tool for sky correction, which does not require this supplementary calibration data, saves a significant amount of valuable telescope time and increases its efficiency. We developed a software package aimed at performing telluric feature corrections on the basis of synthetic absorption spectra.  \end{abstract}

\section{Introduction}
Telescope observing time is very precious. In particular, minimising the calibration overheads for the unavoidable telluric absorption feature correction would lead to a highly increased efficiency as the required observations have to be done directly before or after the actual science frames with the same airmass to capture the same transmission of the Earth's atmosphere. Apart from the loss of science time, this approach additionally introduces severe constraints on the scheduling.

We have developed the software package {\tt molecfit}, which calculates a transmission spectrum on a theoretical basis. Initially, the prototype was developed for estimating the water vapour content of the Earth's atmosphere for scheduling issues of infrared observations with the ESO Very Large Telescope \citep{SME10}. We have further developed this prototype leading to the package {\tt molecfit}, which can now be used for performing telluric absorption feature correction on single and multiple science frames.
\section{The Method}
The tool incorporates a similar approach as \citet{SEI10}, but is more advanced. It is based on the radiative transfer code LNFL/LBLRTM \citep{CLO05}, the line parameter list HITRAN \citep{ROT09}, an atmospheric profile containing information on the chemical composition and temperature of the Earth's atmosphere at the time of observations\footnote{\url{http://www.atm.ox.ac.uk/RFM/atm/}}$^,$\footnote{\url{ http://ready.arl.noaa.gov/gdas1.php}}$^,$\footnote{\url{http://archive.eso.org/asm/ambient-server?site=paranal}}, and the fitting package 'mpfit' by C. Markwardt\footnote{\url{ http://www.physics.wisc.edu/$\sim$craigm/idl/cmpfit.html}}. The fitting algorithm of {\tt molecfit} contains several steps (see Figure~\ref{fig:workflow} and the corresponding user manual \citep{MF13}): (1) scaling the continuum, (2) wavelength and resolution fit, (3) rescaling the continuum, (4) fitting the molecules, (5) joint continuum, wavelength, and resolution fit, and (6) fitting all components (molecules, continuum, wavelength, and resolution). The tool was developed to be instrument independent. It allows the implementation of instrument-specific line kernels for the resolution fit as well as a synthetic kernel composed of boxcar, Gaussian, and Lorentzian components. The best-fit transmission by the radiative transfer code can be used for a telluric absorption correction. The package even has a tool for applying this correction to a set of science observations. Additionally, {\tt molecfit} can be used to determine the water vapour content of the atmosphere for planning infrared observations. In this context, it is also able to fit mid-IR sky emission spectra. The tool has been tested for a selection of spectra taken with different ESO instruments. \articlefigure[width=1\textwidth]{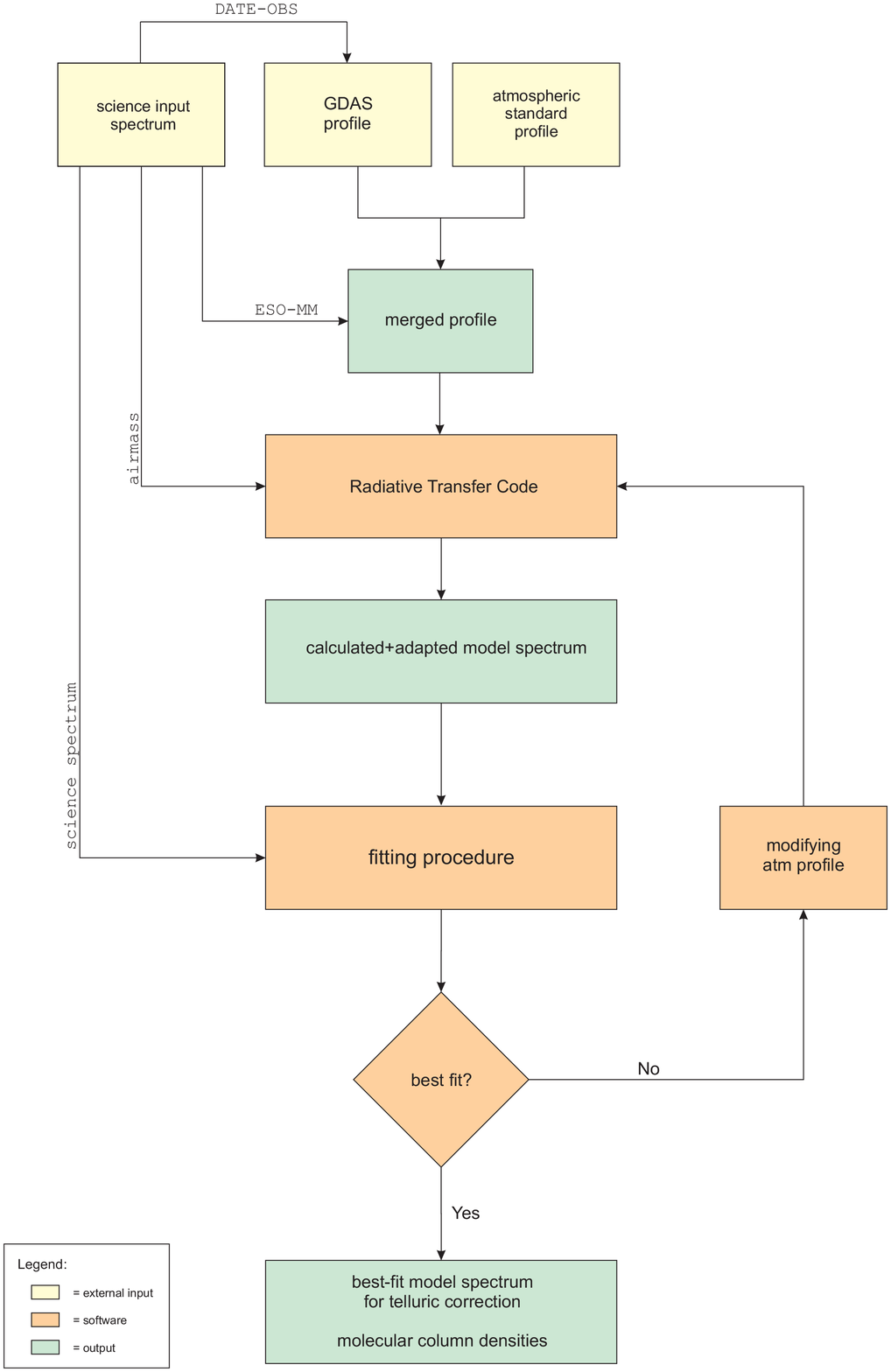}{fig:workflow}{Overview of the {\tt molecfit} workflow}
\section{Results}
We have successfully applied our method to various instruments. Figure~\ref{fig:results} shows a NIR arm spectrum of the galaxy NGC5638 taken with the X-Shooter spectrograph, a medium resolution ECHELLE spectrograph mounted at the ESO VLT covering $0.3$ to $2.5\,\mu$m. We used {\tt molecfit} directly on the science frames without incorporating the corresponding telluric standard star observation. The two regions (I)[$\lambda=1.55$ to $1.65\,\mu$m] and (II) [$\lambda=1.75$ to $2.0\,\mu$m] are defined to show the quality of the corrections with low and high atmospheric absorption, respectively. \articlefigure[width=1\textwidth]{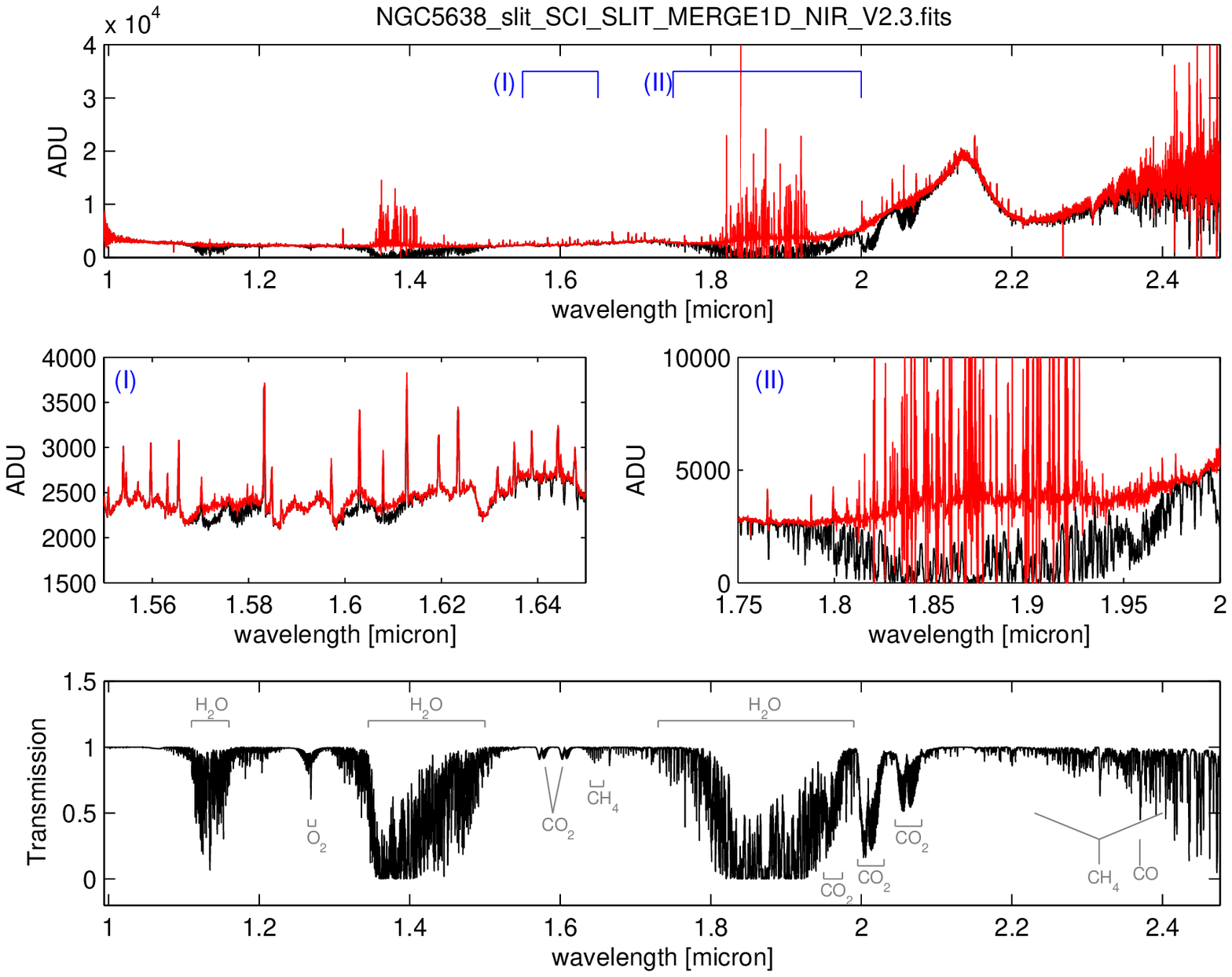}{fig:results}{X-Shooter@ESOVLT example of the telluric absorption feature correction performed by                {\tt molecfit}. The uppermost panel shows the original NIR arm spectrum of the galaxy NGC5638 (black lines) and the corrected one (red lines). Regions (I)[$\lambda=1.55$ to $1.65\,\mu$m] and (II) [$\lambda=1.75$ to $2.0\,\mu$m] show the quality of the telluric feature correction in case of low and high absorption, respectively (middle panels). The lowest panel shows the transmission curve calculated by {\tt molecfit} including an identification of the most prominent molecular features.}
\section{Conclusion}
We obtain a reasonable telluric feature correction in the entire wavelength range of the NIR arm spectrum ($1$ to $2.45\,\mu$m). This is especially demonstrated in the two regions (I) and (II), which shows good results even in wavelength ranges with high molecular absorption by the Earth's atmosphere.  A comparison with the IRAF task {\tt telluric}\footnote{\url{http://iraf.net/irafhelp.php?val=telluric\&help=Help+Page}}, a standard software tool, reveals less residuals and less offsets in the {\tt molecfit} corrected spectra. Thus, performing the telluric absorption feature correction on basis of theoretical transmission curves is a promising approach to save valuable telescope time.

\acknowledgements This study is carried out
in the framework of the Austrian ESO In-Kind project funded
by the Austrian Ministry for Research (BM:wf) under contracts
BMWF-10.490/0009-II/10/2009 and BMWF-10.490/0008-II/3/2011. This
publication is also supported by the Austrian Science Fund (FWF): P26130.

\bibliography{P077}

\begin{thebibliography}{}
\expandafter\ifx\csname natexlab\endcsname\relax\def\natexlab#1{#1}\fi
\expandafter\ifx\csname url\endcsname\relax
  \def\url#1{\texttt{#1}}\fi
\expandafter\ifx\csname urlprefix\endcsname\relax\def\urlprefix{URL }\fi
\providecommand{\eprint}[2][]{\url{#2}}

\bibitem[{{Clough} et~al.(2005){Clough}, {Shephard}, {Mlawer}, {Delamere},
  {Iacono}, {Cady-Pereira}, {Boukabara}, \& {Brown}}]{CLO05}
{Clough}, S.~A., {Shephard}, M.~W., {Mlawer}, E.~J., {Delamere}, J.~S.,
  {Iacono}, M.~J., {Cady-Pereira}, K., {Boukabara}, S., \& {Brown}, P.~D. 2005,
  J. Quant. Spectrosc. Radiat. Transfer, 91, 233

\bibitem[{{Noll} et~al.(2013){Noll}, {Kausch}, {Barden}, {Szyszka}, {Jones}, \&
  {Kimeswenger}}]{MF13}
{Noll}, S., {Kausch}, W., {Barden}, M., {Szyszka}, C., {Jones}, A., \&
  {Kimeswenger}, S. 2013, Molecfit User Manual VLT-MAN-ESO-19550-5772

\bibitem[{{Rothman} et~al.(2009)}]{ROT09}
{Rothman}, L.~S., et~al. 2009, J. Quant. Spectrosc. Radiat. Transfer, 110, 533

\bibitem[{{Seifahrt} et~al.(2010){Seifahrt}, {K{\"a}ufl}, {Z{\"a}ngl}, {Bean},
  {Richter}, \& {Siebenmorgen}}]{SEI10}
{Seifahrt}, A., {K{\"a}ufl}, H.~U., {Z{\"a}ngl}, G., {Bean}, J.~L., {Richter},
  M.~J., \& {Siebenmorgen}, R. 2010, \aap, 524, A11. \eprint{1008.3419}

\bibitem[{{Smette} et~al.(2010){Smette}, {Sana}, \& {Horst}}]{SME10}
{Smette}, A., {Sana}, H., \& {Horst}, H. 2010, Highlights of Astronomy, 15, 533

\end{thebibliography}

\end{document}